# Pathway to Devasthal Astronomical Observatory, ARIES

**Ram Sagar[1] and Gopal-Krishna[2]**

[1] Indian Institute of Astrophysics, Bangalore, 560034 and
Aryabhatta Research Institute of Observational Sciences, Nainital-263001

[2] UM-DAE Centre for Excellence in Basic Sciences, Vidyanagari, Mumbai-400098

**Abstract:** Devasthal observatory, established over a time span of ~ 5 decades, is located in central Himalayan region of Devabhumi in Nainital district of Uttarakhand state, India. Operated and maintained by the Aryabhatta Research Institute of observational sciencES (ARIES), it's location was selected after an extensive site survey. The first measurements of atmospheric seeing and extinctions at Devasthal were carried out during 1997--2001. Since 2010, three optical telescopes of 1.3-m, 3.6-m and 4-m apertures have been successfully installed at Devasthal. Optical and near-infrared observations taken with these telescopes testify to the global competitiveness of Devasthal observatory for astronomical observations. The article chronicles the collaboration with the Tata Institute of Fundamental Research, beginning around 1996, for the purpose of establishing the observatory. A brief overview of the main science results obtained so far, using these facilities, is also presented.

## 1. Introduction

The Aryabhatta Research Institute of Observational Sciences (acronym ARIES) is an autonomous premier research institute under the Department of Science and Technology (DST), Government of India. ARIES is headquartered at Manora peak (longitude = 79° 27′ East; latitude = 29°22′ North; altitude = 1950m), a few Km south of the scenic town of Nainital in the central Himalayan region. Out of its legacy of nearly seven decades, the observatory was nurtured for the first five decades by the Uttar Pradesh (U.P.) and Uttarakhand state governments (Sagar 2022). Devasthal peak, located at an ariel distance of ~22 Km eastward from the Manora peak, is the highest (altitude = 2425 m) mountain peak within ~ 10 Km region. The site is located far from major urban settlements in the region and was chosen after an extensive site survey conducted during 1980–2001 (Sagar et al. 2000; Stalin et al. 2001). The site became operational as an observatory with the commissioning of the 1.3-m Devasthal Fast Optical Telescope (DFOT) in the last quarter of 2010 (Sagar et al. 2011). In March 2007, ARIES awarded contract to the Advanced Mechanical and Optical System (AMOS), a Belgian company, for design, building and installation of the 3.6-m Devasthal Optical Telescope (DOT). The left panel of Fig. 1 shows a snapshot of a design review meeting of the telescope. It was held under the chairmanship of Prof. S. N. Tandon. Others seen participating in the meeting are Prof. P. C. Agrawal, late Mr. S. C. Tapde and Prof. T. P. Prabhu along with AMOS team members and other technical experts. This Indo-Belgian telescope, after its successful installation at Devasthal (Omar et al. 2017; Kumar et al. 2018; Sagar et al. 2019), was technically activated jointly by the Prime Ministers of India, Shri Narendra Modi, and Belgium, Mr. Charles Michel, from Brussels on March 30, 2016 (right panel of Fig. 1). With this, India joined the ranks with the few countries operating such large observing facilities for optical astronomy. This event became an important milestone in the history of Indian optical astronomy as it happened ~ 3 decades after January 6, 1986 when the India's then largest 2.34-m Vainu Bappu Telescope (VBT) was inaugurated at Kavalur in Tamil Nadu near Bengaluru (Bhattacharyya and Rajan 1992).

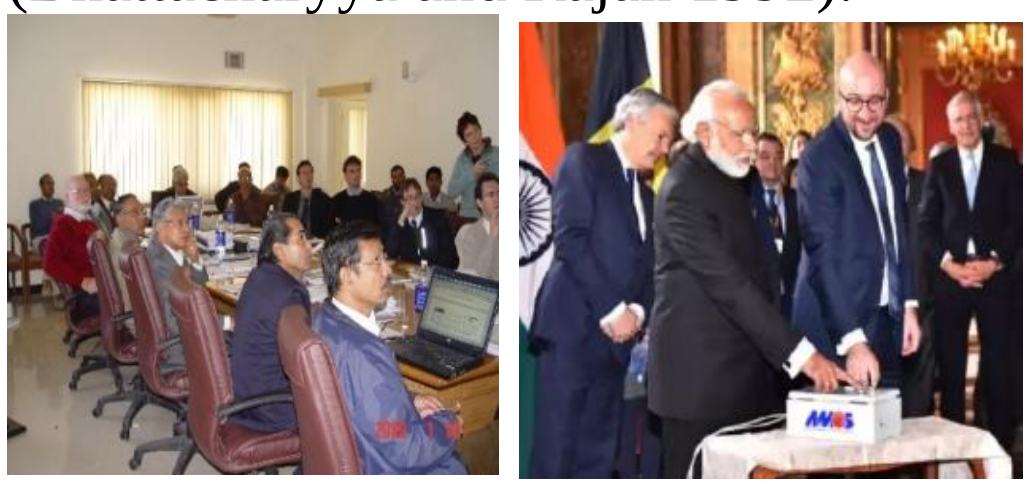

**Fig. 1** The left panel shows a snapshot of the Indo-Belgian 3.6-m DOT design review meeting held on 14-01-2009 at Manora peak, Nainital, while the right panel shows photo of its technical activation on 30-03-2016 from Brussels.

The 4-m international liquid mirror telescope (ILMT), installed at Devasthal (Surdej et al. 2022; Kumar et al. 2022), was inaugurated on March 21, 2023 by His Excellency Mr. Didier Vanderhasselt, ambassador of Belgium in India and Dr. Jitendra Singh, Hon'ble Minister of State (Independent Charge) of the Ministry of Science and Technology & Earth Sciences, in presence of a global astronomical community (Sagar 2023). The left panel of Fig.2 shows a snapshot of the ILMT inaugural function in which dignitaries from both Belgium and India can be seen, including the first author, Prof. Dipankar Banerjee, Director of ARIES and Prof. Jean Surdej, Chief Project investigator of the ILMT from Belgium. An ariel view of the Devasthal observatory is shown in the middle panel of Fig. 2. The bottom right building with the sliding roof is the 1.3-m DFOT. The bottom left rectangular building is the 4 m ILMT. Near the top is seen the 3.6-m DOT dome and the extension buildings. The ancient Devasthal temple and an extended view of the 3.6-m DOT dome are shown in the right panel of Fig. 2.

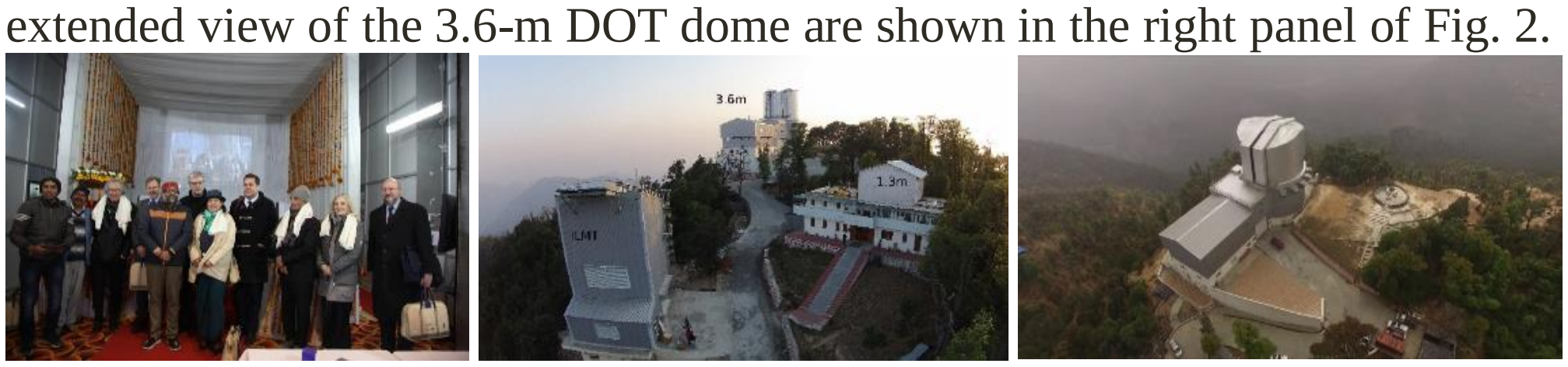

**Fig. 2** From left to right are a snapshot of the ILMT inaugural function, an aerial view of the Devasthal observatory and the ancient Devasthal temple adjacent to the 3.6-m DOT.

The main objective of this article is to provide a sketch of the onerous growth path culminating in the establishment of the Devasthal observatory. Activities during 1976 to 1990 are briefly recounted in the next section. The activities initiated through a major national collaboration are described in Section 3, while the development of infrastructure and the installations of observational facilities are summarized in Sections 4 and 5, respectively. Publications and key science results based on observations taken from Devasthal are briefly outlined in Section 6. The last selection lists our main conclusions followed by the epilogue.

## 2. The initial period (1976-1990), under the Uttar Pradesh state government

During 1976-1977, U.P. government gave an initial in-principal approval to the U.P. State

Observatory (UPSO) for installation of a 4-meter class optical telescope (Sanwal et al. 2018; Sagar 2022). Following the approval, the project funding began in the financial year 1978-1979. The site survey work was launched in1980 and in this connection an exploratory trip to USA and UK was also undertaken by an astronomer of the UPSO. In January 1981, a Project Technical Board (PTB) was constituted for monitoring and implementation of the project. The PTB recommended UPSO to collaborate with Bhabha Atomic Research Centre (BARC), Mumbai for preparing a Project Concept Report (PCR). Accordingly, a team consisting of members from both institutions visited Australia, UK and USA during 1983 and completed the PCR in 1984. Prof. M.G.K. Menon, member, Central Planning Commission (CPC), constituted a national committee which recommended a 2.34-m, instead of 4-m telescope with a funding of Rupees 600 lakhs. However, U.P. government did not concur and requested the CPC for approval of a 4-m class telescope project. Consequently, a team of CPC under Chairmanship of Prof. Menon visited UPSO, Nainital in June 1986. After an extensive discussion with the UPSO, the CPC team suggested a 4-m class telescope project as a national facility, by involving other Indian institutions. Accordingly, in February 1987, meeting of a national expert committee was convened at BARC, Mumbai. This committee formally recommended the 4-m class telescope project to function as a national facility. In August 1987, the CPC under the Chairmanship of Prof. Menon, accepted this recommendation and advised preparation of a Detailed Project Report (DPR) in cooperation with the Department of Atomic Energy (DAE). During 1988-89, the DST organised several meetings of the Indian astronomy community to prioritise proposals for national observing facilities in the area of astronomy during the $8^{th}$ five-year plan (1990-1995). One of the major recommendations of these meetings was the installation of a large national optical telescope at a suitable site in India before the end of the $20^{th}$ Century (Anupama et al. 2022). Unfortunately, the telescope project activities could not be pursued after April 1990 due to lack of funding from the U.P. Government (Sanwal et al. 2018).

### *2.1 Site survey activities during 1980 - 1990*

Manora peak is not well suited for installation of 3 to 4-m class optical telescopes, mainly due to degraded atmospheric seeing (Sagar 2022). Therefore, a site survey was initiated in 1980-81 in the Kumaon and Garhwal regions of the then U.P. (now Uttarakhand) state. For this, Survey of India's contour maps (1:50000) of these regions were studied and a total of six reconnaissance trips were undertaken to 36 sites during 1981-82. Based on altitude of the site and its viewing obstructions due to nearby hills, terrain of the surrounding regions and logistic factors, like availability of reasonably flat patch of land, the availability of water source, distance from an existing 6 m wide metalled road and the likelihood of growing light pollution due to nearby towns and cities in the foreseeable future, a total of five sites namely Gananath, Mornaula, Devasthal, Jaurasi and Chaukori (all having altitude > 2 km) were identified for further investigations. The meteorological instruments installed at these stations were thermograph, hygrograph, barograph, sunshine recorder, rain gauge, snow gauge, and wind speed and direction recorder (Sagar et al. 2000; Sanwal et al. 2018). Left panel of Fig. 3 shows one such instrument installed at Devasthal. These meteorological observations were carried out at all the five sites, for a decade.

For measuring the atmospheric seeing, a key parameter for locating modern optical telescopes, a 25-cm polar star trail telescope (right panel of Fig.3) was used. As the name indicates, such telescopes always point toward the North pole star. Star trails were recorded on photographic films using a Kodak SLR, 35-mm camera. Such observations were carried out at two most promising sites, namely Devasthal and Gananath. The profile of the star trail was measured at a number of points using a microdensitometer. The quantitative estimate showed that most of the time the seeing at Devasthal was better than that at Gananath. The stability of night time temperature, relative humidity, wind speed and direction, and logistics also favoured Devasthal site (Sanwal et al. 2018).

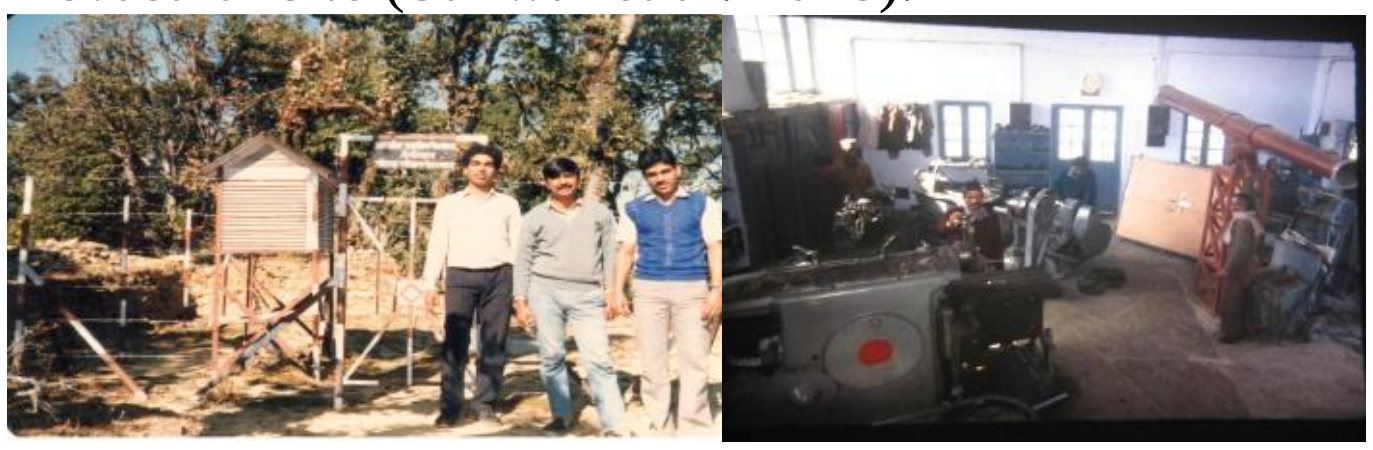

**Fig. 3** The left panel shows a meteorological station. The right panel displays 25-cm polar star trail telescope in the mechanical workshop. It was used for seeing measurements at Devasthal.

## 3. Renaissance of telescope project activities at Devasthal site (1996-2003)

In July 1996, the first author moved from Indian Institute of Astrophysics (IIA), Bangalore (now Bengaluru) to the then UPSO, Nainital as its director. Around that time, the Tata Institute of Fundamental Research (TIFR), Mumbai and its Pune branch, the National Centre for Radio Astrophysics (NCRA) expressed interest in establishing a medium to large size optical telescope in collaboration with the UPSO. In September 1996, the Secretary, DST, U.P. state government, Mr. Prabhat C. Chaturvedi visited the Manora peak and Devasthal sites of UPSO. During the visit, he briefed media about the future plan for Devasthal observatory. In the following years, he played a key role in reviving activities of the large telescope project. So, after a gap of ~ 6 years, characterization of Devasthal site was resumed in collaboration with IIA and TIFR. All these initiatives paved the way for the eventual installation of three telescopes, namely the 1.3-m DFOT, the 3.6-m DOT and the 4-m ILMT at Devasthal. All three are currently operational.

The rationale behind the collaboration between UPSO and TIFR was that although TIFR had made major strides in setting up observational facilities in radio, infrared, X-ray and γ-ray branches of astronomy, a need was strongly felt for backing up these multi-wavelength observational facilities with an easy-to-access modern optical telescope of 3-m class. Therefore, when the move of the first author to UPSO as its director became imminent in January 1996, some TIFR astronomers saw emergence of opportunity to plug the long-standing critical gap in TIFR's astronomy capability. The plan crystallised in the conception of a joint project between UPSO and TIFR, for setting up a medium/large aperture (> 2-m) optical telescope at a site to be provided and managed by UPSO. The staff of UPSO already possessed significant expertise in optics, site surveying and telescope maintenance and operation, whereas TIFR had acquired a formidable engineering expertise (structural, mechanical, electronics and software development) in several areas of multi-wavelength astronomy. This expertise had been gained through indigenous building of a large radio telescopes at Ooty in 1960s and the world-class Giant Metrewave Radio Telescope (GMRT) at Khodad (near Pune) during 1990s (Swarup 2021), as well as the balloon-based infrared/X-ray telescopes, later culminating in the successful launch of the first Indian space observatory (AstroSat) for UV/X-ray astronomy, etc (Agrawal 2017). The potential location for the proposed optical telescope was the logistically and optically suitable site of Devasthal.

As the first step, a note was prepared by the second author, which was taken up for discussion in the meeting of the TIFR Group Committee II (GC II) held on April 2, 1996 to finalise the IX 5-year plan (1996--2001) proposals. Several members of the GC II agreed to participate in the proposed optical telescope project, in particular, Prof. P. C. Agrawal, Prof. T. N. Rengarajan who emphatically endorsed the proposal. GC II nominated Prof. Gopal Krishna and Prof. S. K. Ghosh as the PIs from TIFR side, with major involvement of late Mr. S. C. Tapde (TIFR) in design and construction of the facility, based on his experience as the Project Manager of the 2.34-m VBT, and thereafter of the ongoing GMRT project. On April 25, 1996, Chairman of the Budget & Planning Committee of TIFR, Prof. G. Govil, via a letter addressed to Profs. Gopal Krishna and T.N. Rengarajan, informed that TIFR's Internal Working Group had recommended a tentative allocation of Rs. 18.00 Crore for the optical telescope project and sought a detailed project proposal with year-wise breakup of the budgetary projection along with justification, for the purpose of DAE's IX 5-year Plan document to be submitted to the CPC. This matter was also discussed in the faculty meeting of NCRA-TIFR (Pune) on May 11, 1996, which strongly endorsed the optical telescope project. The needed year-wise break-up of the budgetary projections, together with the manpower requirement, was communicated by Prof. Gopal Krishna to the TIFR Dean, Prof. G. Govil in his letter dated May 14, 1996. On the same day, he made a detailed presentation at TIFR, Mumbai, before the Internal Working Group, about the proposed optical telescope joint project between TIFR and UPSO. In addition to highlighting the criticality of optical back-up to TIFR's ongoing multi-wavelength astronomy operations, he also underscored the importance of the geographical location of Devasthal site which filled the wide (northern) longitudinal gap between the Canary Islands and Australia where major optical telescopes were then stationed.

A communication dated August 27, 1996, from TIFR Director, Prof. V. Singh to the Director of UPSO formally conveyed TIFR's interest in setting-up a 2-metre class optical telescope at Devasthal site, jointly with UPSO on cost-sharing basis, as part of the IX 5-year Plan. In response to that letter, Shri P. C. Chaturvedi, Secretary, DST (Govt. of U.P.) and the Director, UPSO visited TIFR and participated in a meeting held on January 2, 1997 which was chaired by Prof. V. Singh (Director, TIFR). The other participants included Profs. G. Govil, P.C. Agrawal, T. N. Rengarajan, S.K. Ghosh, K.P. Singh and Gopal Krishna. During the meeting Shri Chaturvedi reaffirmed the willingness of U.P. government to set up a 'modern 2-metre class optical telescope' at Devasthal, as a collaborative project between UPSO and TIFR with a total outlay of Rs. 30 Cr, on a 50:50 cost sharing basis. Shri Chaturvedi also expressed the optimism that with TIFR's collaboration, the project should go through smoothly. Thereafter, on February 4, 1997, TIFR Director communicated the outcome of the above meeting to Shri A. Dasgupta, Jt. Secretary (R&D) of DAE. On March 26, 1997, Shri Dasgupta wrote that DAE had no objection to this project being included in TIFR's IX 5-year Plan document. Under this plan proposal (9P-1001), in order to facilitate purchase of equipment for the site chracterisation at Devasthal, the Internal Working Group of TIFR made a formal allocation of Rs. 10 lakhs for the financial year 1997-98, as communicated to the Director UPSO in a letter dated July 10, 1997 by Prof. P. C. Agrawal, incumbent Dean, Physics Faculty, TIFR. However, this sanctioned grant remained unutilized. A flavour and perspective of these initial endeavours towards the genesis of the TIFR-UPSO collaboration can be gleaned from the few documents reproduced in Annexure I.

A DPR, prepared jointly by UPSO and TIFR for setting up a 3-metre class modern optical telescope at Devasthal, was approved by the CPC during 1998-1999 and a Memorandum of Understanding (MoU) was signed on May 8, 2000 between TIFR and UPSO which led to the constitution of a Project Steering Committee (PSC) consisting of members from both institutions with Director, TIFR as its Chairman and the Chief Secretary, U.P. Government as co-Chairman (Annexure II). Left panel of Fig. 4 shows a snapshot of the first PSC meeting chaired by Prof. S. S. Jha (seen in the middle). Also seen are Prof. Ram Sagar, Dr. S. D. Sinvhal and Prof. R. P. Verma seated on one side and Prof. Gopal Krishna, Prof. N. K. Sanyal and two officers from U.P. Government seated on the other side. On June 29, 2000, several PSC members visited Devasthal despite formidable logistical challenges in accessing the site, further aggravated due to the onset of monsoon. Both institutions agreed to allocate necessary funds for the telescope project. The first transfer of funds to the tune of Rs. 3.66 Cr was made by the U.P. Govt. on March 27, 1999 (Appendix III). However, at that point a major event struck with the creation of a new state of Uttarakhand carved out of U.P. on November 9, 2000 and immediately UPSO came within the ambit of the new state and rechristened as "*State Observatory*". On 12 February 2001, the Uttarakhand State Government approached DST (Government of India) for taking over the State Observatory. This development was discussed during the 2$^{nd}$ meeting of the PSC, held on February 22, 2001 at Indian National Science Academy headquarter in New Delhi. Since the Government of Uttarakhand expressed its inability to fund the telescope project due to its prevailing financial constraints, this collaborative project with TIFR had to be shelved. Through a Cabinet decision of the Government of India, taken on January 7, 2004, the state government funded 50 years old observatory mutated into ARIES on March 22, 2004 (Sagar 2022). Thereafter, activities related to development of Devasthal observatory regained momentum.

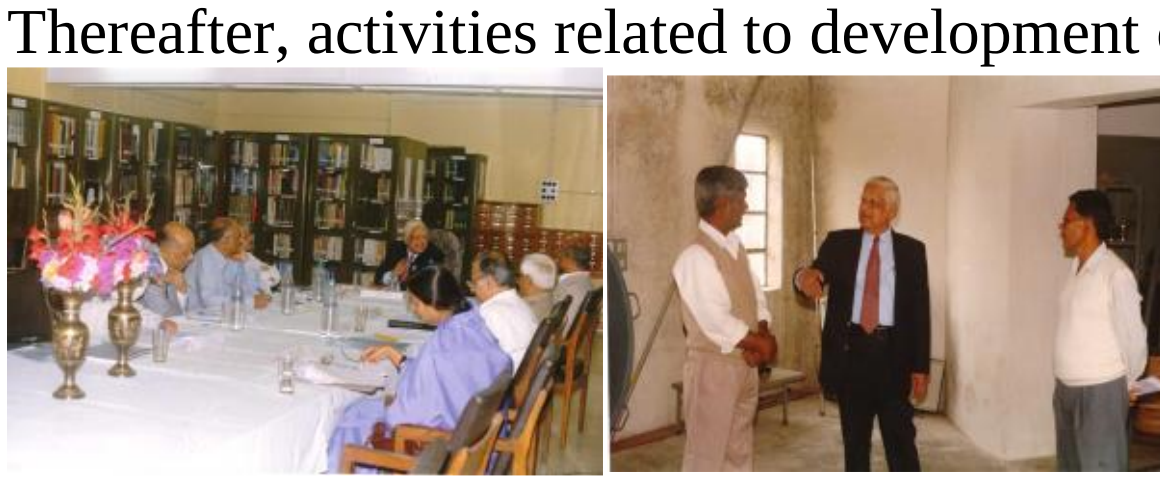

**Fig. 4** The left panel shows a snapshot of the first PSC meeting held at Manora peak on June 28, 2000. In the right panel, Dr. Vijay Mohan and Dr. B. B. Sanwal are seen in conversation with Prof. S.S. Jha, director, TIFR and chairman, PSC.

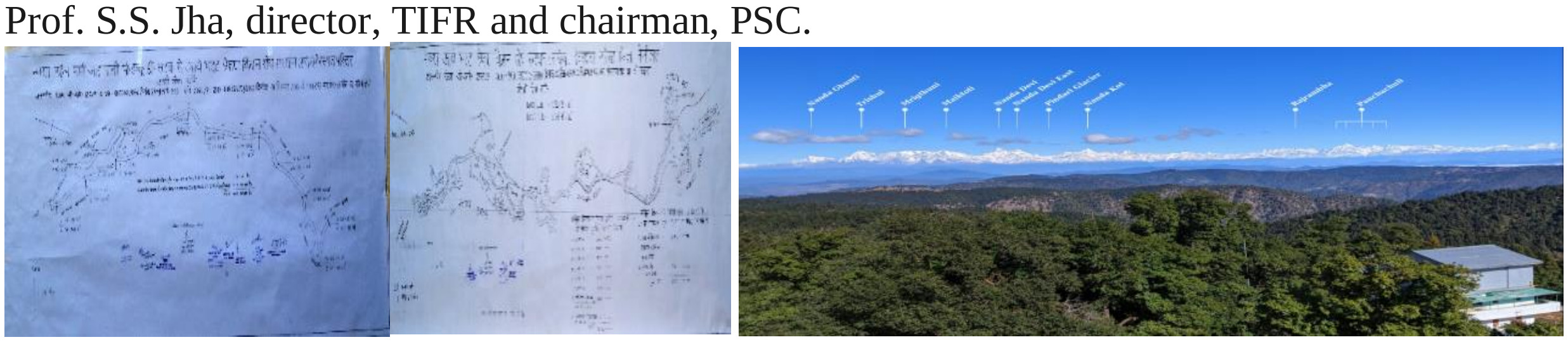

**Fig. 5** Maps of the land acquired by ARIES for Devasthal observatory are shown in the left and middle panels. They are located in North and South Gola forest-regions of Kumaon. The right panel shows a wide-angle view of the western Indian Himalayan snow peaks. This splendour was captured on 4-12-2021 from top of the 3.6-m DOT dome (courtesy: Dr. Brajesh Kumar).

The infrastructure and connecting roads (left and middle panels of Fig. 5) of Devasthal observatory are spread across a region of over 5 ha located in 5 “Van Panchayats” (Sagar et al. 2019). The right panel of Fig. 5 depicts a breath-taking vista of the tallest Indian mountains located in the western Himalayan range. This photo was taken around noon of December 4, 2021 from top of the 3.6-m DOT dome. A few cloud patches rising from snow peak are clearly seen under the blue transparent sky. The 1.3-m DFOT building is in the bottom right while prominent snow-clad peaks seen, from left to right, are Nanda Ghunti (6309 m), Trishul (7120 m), Mrigthuni (6855 m), Maiktoli (6803 m), Nanda Devi (7816 m), Nanda Devi East (7434 m), Nanda Kot (6861 m), Rajrambha (6537 m) and Panchachuli (6312 to 6904 m). First peak is located in the Garhwal region while all others are in the Kumaon region. The Panchachuli peaks are a group of five snow-capped Himalayan peaks lying at the end of the eastern Kumaon region, near Munsiyari, in Pithoragarh district of Uttarakhand state. Well known Pindari Glacier (3, 353 m), located between Nanda Devi East and Nanda Kot, can also be clearly seen.

### *3.1 Seeing and extinction measurements at Devasthal during 1997 to 2001*

As mentioned earlier, UPSO revived the Devasthal site survey activities in collaboration with IIA and TIFR in 1996. The IIA provided one-time financial grant of Rs. 6 lakhs while UPSO and TIFR jointly started building the essential infrastructure at Devasthal, including the laying down of a upto 3- Km long footpath to the summit. The existing 38-cm and 52-cm optical telescopes of UPSO were relocated from Manora peak to Devasthal site and used for the first time, for measuring atmospheric extinction (Mohan et al. 1999) and seeing (Sagar et al. 2000; Stalin et al. 2001; Sagar et al. 2019). Fig. 6 shows a photograph of the 52-cm telescope installed near the base camp of Devasthal. Standing in front of the telescope under the blue sky are, from left to right, Prof. S. K. Sateesh, Dr. Wahab Uddin, Profs. S. K. Ghosh and Ram Sagar, Mr. P. S. Bhaisora, Profs. P. C. Agrawal, P. Venkatkrishanan, R. Srianand and Gopal-Krishna, Mr. K. S. Rawat, Dr. B. B. Sanwal, Prof. T. P. Prabhu and late Dr. A. K. Pandey.

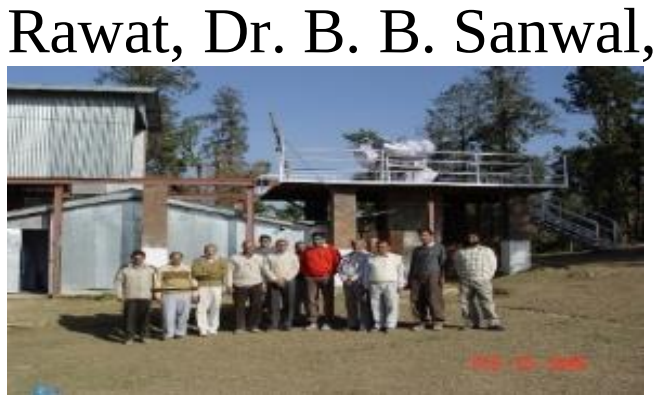

**Fig. 6** A snapshot taken on 25 February 2006 shows the 52-cm optical telescope installed at the base camp of Devasthal observatory site.

## 4. Development of the infrastructure at Devasthal

For installing modern optical telescopes at Devasthal, infrastructure development was started immediately after formation of ARIES in March 2004. Forest land, left and middle panels of Fig. 5, transferred to ARIES was used for constructing approach roads linking the main PWD road to the base camp and to the telescope sites located near Devasthal peak. A brief account of these activities is given below.

### *4.1 The approach road from Jarapani to Devasthal*

The foundation stone of this ~ 3 Km long and ~ 4 m wide approach road was laid on October 27, 2005 by late Dr. S. D. Sinvhal, a Governing Council member of ARIES. Fig. 7 shows two snapshots of this watershed event. Left panel shows foundation stone laying ceremony. In the right panel, seated on the dais are, from left to right, Mr. R. L. Vishwkarma, Dr. K. Sinha, Prof. Kavita Pandey, Dr. M.C. Pande, Dr. S.D. Sinvhal, Mr. Sher Singh Naulia, Prof. Ram Sagar and Dr. B.B. Sanwal. Several villagers can be seen watching the function from the roof of an adjoining house in the background. That crucial road connects Jadapani village located along the main PWD road to Devasthal peak. Earth cutting work for this road was completed by May 2006. After its stabilization during the rainy seasons (June to September) of 2006 and 2007, laying of the inter and top coat of the approach road was completed by December 2007. This road is open for public and serves the important purpose of transporting heavy equipment and other materials not only to Devasthal but is also used for visiting the Devasthal temple by public and the residents. However, routine maintenance of the road is the responsibility of ARIES.

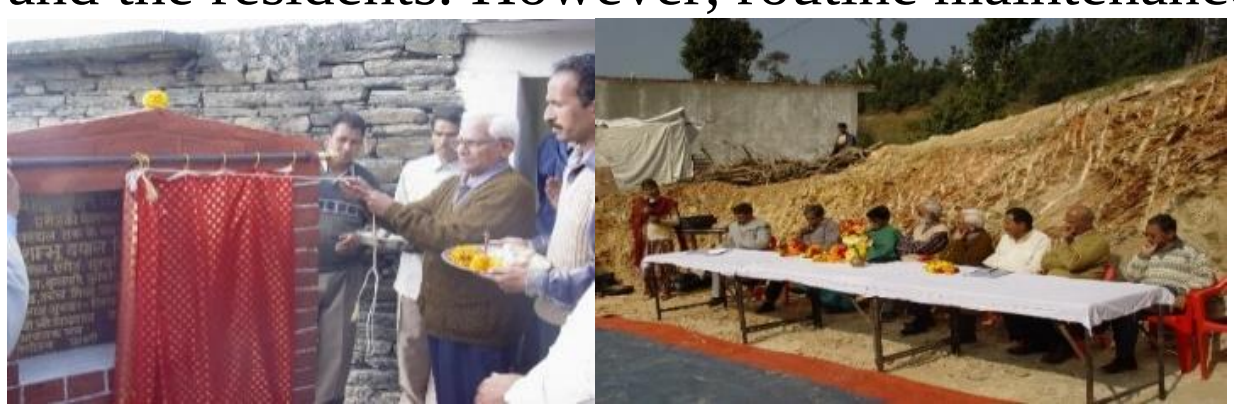

**Fig. 7** Snapshots of the foundation stone laying function held on October 27, 2005.

### *4.2 The base camp and infrastructural support facilities*

Measurements published by Stalin et al. (2001) indicated that atmospheric seeing is the best near Devasthal peak. This was also expected from the topography of the region (Sagar et al. 2019). Therefore, the area near Devasthal peak was earmarked for installation of optical telescopes while two other nearby locations were identified for construction of the required support facilities such as a guest house and an electrical substation, etc. A small flat area located ~ 500 m away from the peak was used for installing an electrical substation and generators etc. to meet the electricity requirements. A base camp was set up on a flat patch of land, nearly 120 m below the Devasthal peak and ~ 1.5 km away from it towards the Jadapani village. This area was used to set up other infrastructural support facilities as well. Growth of these facilities is therefore expected to have bare minimum contribution to light pollution and to deterioration of natural atmospheric seeing near the telescopes. Although activities related to installation of the 1.3-m DFOT were already in full swing, no accommodation was available at the site. Hence, during the summer of 2008, two pre-fabricated huts were installed at the base camp. This became the first reasonably furnished single-room accommodation for visitors coming to Devasthal. Ms. Sheila Sangwan, then Financial Advisor and Joint Secretary, DST, was the first senior visitor to stay overnight in this hut during her official visit to ARIES in May 2010. The *Rest Room*, a modern guest house subsequently set up at the base camp, has five rooms with attached bath. Its dining area has a capacity for ~ 15 persons. This building was inaugurated by Prof. G. Srinivasan on May 9, 2011. In the photo (Fig. 8), Mohit Joshi, Jagdish Chand Tiwari (Pujari) and Ram Sagar are seen on left side while G. Srinivasan, T. Bhattacharyya and Om Prakash are towards right side. For taking rest after the observations, two rooms in the 1.3-m DFOT building on the ground floor and another two rooms in the basement of the ILMT building were also constructed.

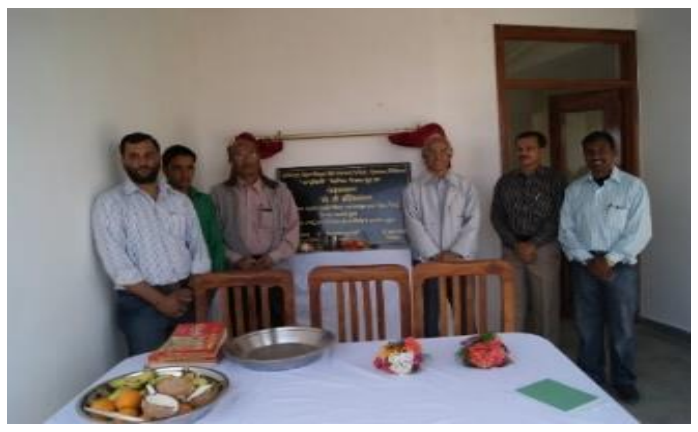

**Fig. 8** Photo of the inaugural function of the Devasthal *Rest Room* on May 9, 2011.

In order to ensure availability of water for daily use at Devasthal observatory, digging of a bore well was successfully carried out and the required number of water tanks were constructed at appropriate heights and locations. Laying of a pipeline including the installation of a submersible pump were also completed. All these activities were carried out during 2006 to 2008. Tap water is now available in all buildings at Devasthal including the 1.3-m DFOT, 3.6-m DOT and 4-m ILMT buildings. In coordination with the Uttaranchal Electricity Board, a dedicated feeder for supplying electrical power was commissioned and two substations with capacities of 150 KVA and 65 KVA were set up at Devasthal. Activities related to electrical connections to all the telescope facilities and buildings were planned and implemented by the ARIES as per the requirement.

During 2004--2005, for online transfer of the digital data from the Devasthal observing facilities to the Manora peak campus of ARIES, a 2.4 GHz microwave link of total usable bandwidth of 14 Mbps was installed. As the data flow increased, the microwave link was replaced with a BSNL optical-fibre link having a 30 Mbps bandwidth. In addition to this, two microwave connections (PTP Link) were also made available for data and voice transfer between Devasthal and Manora peak campuses of ARIES.

## 5. Observational facilities at Devasthal

Observational facilities at an astronomical observatory are primarily used to take observations of celestial objects. For this, back-end instruments are mounted at the focal plane of the telescope. It is the prime focus in case of the 4-m ILMT and Ritchey–Chre´tien Cassegrain (RC) focus for both 1.3-m DFOT and 3.6-m DOT. The 4K x 4K CCD imagers are used at the RC focus of the 1.3-m DFOT (Joshi et al. 2022) and at the prime focus of the 4-m ILMT (Kumar et al. 2022). Three back-end instruments (one on the main axial port and two on the side ports) can be mounted simultaneously at the RC focus of the 3.6-m DOT (Kumar et al. 2018; Sagar et al. 2019). The UPSO-TIFR collaboration, initiated about 3 decades ago, has continued to progress in the area of back-end instrumentation, particularly for near-infrared (NIR) observations with the 3.6-m DOT. The TIRCAM2 (TIFR NIR Imaging Camera-II) is permanently mounted on a side port while TANSPEC (TIFR-ARIES NIR Spectrometer) can be mounted on the axial port of the 3.6-m DOT (Sharma et al. 2022a; Ojha et al. 2018). For observations in the optical band, both 4K x 4K CCD IMAGER and ARIES-Devasthal Faint Object Spectrograph and Camera (ADFOSC) developed and built by ARIES, are mounted on the axial port of the 3.6-m DOT (Omar et al. 2019a; Kumar et al. 2022; Dimple et al. 2023). All these instruments are playing a key role in realising the scientific potential of the Devasthal observatory which is in regular operation as a national facility. Currently all the back-end instruments are available for observations to all astronomers with Indian or foreign affiliation, through peer reviewed scientific proposal(s), which are invited at regular intervals through public notifications.

## 6. A snapshot of the scientific output based on observations taken from Devasthal

After testing and characterization of the back-end instruments, the observing facilities of Devasthal were released for observations. The observing proposals mostly covered topics related to photometric and spectroscopic studies of galactic star-forming regions, star clusters, Active Galactic Nuclei (AGN) including quasars/blazars, supernovae, optical transient events, distant galaxies, etc. Both 1.3-m DFOT and 3.6-m DOT are in regular use for the astronomical observations while the 4-m ILMT has just entered the operational phase (Surdej et al. 2022, 2023). Observations taken with these facilities have been reported in over 200 publications and have led to about two dozen PhD theses so far. Table 1 gives their distribution with time, for different research areas. About 70 % of the publications are in standard peer-reviewed journals while the remaining are published elsewhere, including GCN circulars. Two of them have appeared in 'Nature', a journal of very high impact factor, and another 6 papers have appeared as Letters in high-impact top-rung international journals published from U.K. and U.S.A.. Publications until 2010 are mostly related to telescope projects and characterisation of Devasthal site. Observations taken with the 1.3-m DFOT have contributed to publications from the year 2011 at the rate of 4 to 5 per year. Since 2016, observations taken with the both 1.3-m DFOT and 3.6-m DOT have contributed to the publications. Consequently, the annual publications rate has increased significantly from ~ 5 to ~ 24. Similar growth is also witnessed in the PhD theses using observations made at Devasthal Astronomical observatory.

**Table 1.** The growth of publications and PhD theses based on observations taken from Devasthal. Data are taken from website of ARIES. During the period mentioned in the first column, number of publications in the fields of Galactic, Extragalactic, Transient astronomy and others (site characterization and observing facilities etc.) and their total number are listed in columns 2, 3, 4, 5 and 6, respectively. The last column gives the number of PhD theses awarded.

| Period in years | Galactic astronomy | Extragalactic astronomy | Transient astronomy | Others | Total number | PhD theses |
|---|---|---|---|---|---|---|
| 1999-2010 | 0 | 0 | 0 | 10 | 10 | 0 |
| 2011-2016 | 4 | 15 | 6 | 25 | 50 | 5 |
| 2017-2022 | 39 | 39 | 41 | 24 | 143 | 18 |

Key scientific results and the potential of **Devasthal observing facilities are summarized below.**

- Both optical and NIR observations taken with the 3.6-m DOT reveal that sub-arcsec quality images can be obtained for a significant fraction of photometric nights. They also indicate that performance of the telescope is at par with its peers located elsewhere in the world (Omar et al. 2017; Kumar et al. 2018; Sagar et al. 2019, 2020, 2022). Consequently, the facilities at Devasthal have led a good number of national and international collaborations including the Indo-Belgian techno-scientific collaboration termed BINA (Joshi and De Cat 2019; Sagar 2023). TIFR contribution in TANSPEC and TIRCAM2 NIR instruments is an excellent example of intra-national collaboration (Sharma et al. 2022a, b; 2023 and references therein). TIRCAM2 is now permanently mounted on the 3.6-m DOT and is open to all users. Since March 2016, lunar occultation observations have been performed using both 1.3-m DFOT and 3.6-m DOT telescopes. These milliarcsecond accuracy NIR observations have yielded angular diameters of many cool stars and resolved close binaries, providing valuable insight to the formation and evolution of these objects (Joshi et al. 2022). Similarly, precise observations of a stellar occultation by Pluto, taken on June 6, 2020 with the 1.3-m DFOT and 3.6-m DOT telescopes, were used to place important constraints on the evolution of Pluto's atmosphere (Sicardy et al. 2021).

- Star clusters are ideal laboratories for studying stellar variability, formation of stars and stellar evolution. Several open star clusters and a few globular clusters located in our Milky-way galaxy have been observed at optical and NIR wavelengths using Devasthal observing facilities. These observations combined with GAIA kinematical and other archival data have led to the studies of star formation efficiency, age distribution, mass function and luminosity function in open star clusters (Panwar et al. 2017, 2022; Pandey et al. 2020; Maurya et al. 2021; Sagar et al. 2022; Sharma et al. 2023 and references therein). Spatial structure and interstellar extinction in young open star clusters have also been studied. The slope of the initial mass function above 1 solar mass for young star clusters (age < 100 Myr) is found to be universal, consistent with the Salpeter value within uncertainties of its observational determination. Stellar variability studies in star clusters have led to discovery of many new variables (Lata et al. 2021; Sagar et al. 2022; Maurya et al. 2023).
- Gamma Ray Bursts (GRBs) are relatively short-duration (<1 second to several minutes) most energetic events observed in the Universe since the Big Bang. The first Indian optical observations of a GRB afterglow (GRB 990123) were made at ARIES (Sagar 2022). Since then, over 50 GRB afterglows, supernovae and novae have been successfully observed with the telescopes of ARIES. Deep photometric and low-resolution spectroscopic observations taken from Devasthal were combined with the published ones, including those at other wavelengths, for probing the nature of GRB afterglows, supernovae and novae as part of transient astronomy research (Ailawadhi et al. 2023; Gupta et al. 2022 and references therein). The optical light-curves, spectral energy distributions and estimated energetics of these bursts are used for placing observational constraints on the progenitors of GRBs, supernovae and novae (Pandey et al. 2021 and references therein). They support the core collapse model for progenitors of long-duration GRBs and place observational constraints on the popular progenitor's models for short-duration GRBs (Troja et al. 2022). Observations of GRB afterglows taken with Devasthal observing facilities have contributed over 45 publications in high impact journals, in addition to a few PhD theses.
- Optical variability of powerful AGN has been studied at ARIES since the late 1990's. For this, sensitive observations have been performed for detecting intra-night optical variability down to ~ 1 % level. The collaboration between TIFR and UPSO/ARIES has continued and proliferated in the area of rapid optical variability of AGN. This intriguing type of variability, termed "Intra-night optical variability" (INOV: Gopal-Krishna et al. 2003; 1995) has evolved into a fairly major branch of AGN research world-wide, since it allows probing the physical conditions of powerful AGN on micro-arcsecond scale which is not amenable to direct imaging at any waveband. Until mid-1990s, the young science of INOV was essentially confined to blazars (dominated by relativistically beamed nuclear jets) and low-luminosity AGN, namely Seyfert galaxies. Extending this to several other prominent classes of AGN, including the radio-quiet quasars (RQQs) was taken as a priority task at ARIES. In order to carve a niche in this field, a long-term program was conceived and spearheaded by the second author (GK) using the 104-cm Sampurnanand telescope located within the ARIES main campus at Manora peak. Subsequently, it was extended to the 1.3-m DFOT when this new 'work horse' became operational in the year 2010. Other than the present authors, active participants in this particular program include Paul J. Wiita, Hum Chand and Ravi Joshi, as well as 7 ARIES research scholars, all of whom have completed their PhD theses. The prominent AGN classes whose INOV characteristics were thus systematically established, for the first time and with good statistical significance are: (1) radio-quiet quasars, (2) radio-intermediate quasars, (3) core-dominated radio quasars, (4) lobe-dominated radio quasars, (5) TeV detected blazars, (6) radio-quiet weak-emission-line quasars, (7) narrow-line Seyfert1 galaxies, (8) broad-absorption line quasars, (9) high-redshift blazars (actually monitored in UV band in the rest-frame due to their high redshifts) and (10) low-mass quasars. At a high significance level, superluminal motion and optical polarisation were demonstrated to be the key markers of INOV, rather than radio loudness or flat radio spectrum, as claimed by many authors previously. The vast body of novel results obtained under this niche project, entirely using small-size Indian telescopes, have been reported in over 40 publications (including 28 in peer-reviewed international journals), altogether receiving over 900 citations (Gopal-Krishna and Wiita 2018; Chand et al. 2018; Ojha et al. 2020; Mishra et al. 2021; Gopal-Krishna et al. 2023 and references therein). Publications using the Devasthal facilities, under this program, include 5 Letters to the leading British journal (MNRAS) during 2019-23. Around 2008, another major project on AGN variability (including INOV) was launched at ARIES, led by Alok C. Gupta. It has largely remained focused on blazars. Unlike the afore-mentioned first INOV program at UPSO/ARIES, this second program has gradually evolved to become part of an international network of telescopes and astronomers spread across many countries, including the 'Whole Earth Blazar Telescope (WEBT)' (Gupta et al. 2016; Raiteri et al. 2021; Dhiman et al. 2023 and references therein). Main topics investigated under this program are: (1) optical colour variations associated with brightness changes, (2) multi-band variability correlations, (3) quasi-periodicity in blazar light-curves, (4) variations of the 'spectral-energy-distributions'. This project has led to over 35 publications in peer-reviewed international journals and to PhD theses of 4 research scholars of ARIES.

- Devasthal observing facilities have also been used for optical observations of different type of galaxies. The Wolf–Rayet dwarf galaxy Mrk 996 was the first object observed by Jaiswal and Omar (2013) at optical wavelengths. Omar et al. (2019b) detected TGSS J1054+5832, a GMRT-detected high-redshift ($z = 4.8 \pm 2$) steep-spectrum radio galaxy, at optical wavelengths, using ADFOSC mounted on the 3.6-m DOT. These observations highlight the importance and potential of this telescope for detections of faint galaxies.

## 7. Summary

Devasthal, located in Nainital district of Uttarakhand state, was identified as a potential astronomical site based on an extensive site survey conducted during 1980-1990. Recent studies by Ningombam et al. (2021) and Priyatikanto et al. (2023) indicate that it continues to be a globally competitive astronomical observatory and can be rated as a very good site in the Asian region. A comparison of atmospheric seeing and extinction measurements carried out ~ 2 decades ago with those estimated from recent observations taken with the 3.6-m DOT shows that sky conditions at Devasthal have not deteriorated (Sagar et al. 2019, 2020; Joshi et al. 2022). This confirms that the abundant care taken during infrastructure development and construction of the telescope buildings has paid rich dividends. It can, therefore, be asserted that the observing facilities located at Devasthal have the potential of providing internationally competitive and scientifically valuable optical and NIR observations in a number of frontline areas of Galactic and extra-galactic astrophysical research, including the vital optical follow-up of GMRT and AstroSat sources. Devasthal observatory has also proved to be a fertile ground for a number of collaborations, e.g., between UPSO/ARIES and TIFR. Last, but certainly not the least, the geographical location of the Devasthal observatory offers a special advantage for the time domain and multi-wavelength astrophysical studies (Sagar et al. 2019, 2020, 2022).

## Epilogue

The purpose of this historical note is to chronicle the motivation, initiatives and events that laid the foundation for the emergence of the astronomical observatory at Devasthal, currently hosting India's largest sized 4-m class optical telescopes. That sequence of activities during the UPSO era also imparted the momentum to the prospects of transition of UPSO into ARIES funded by the government of India. As they say, gaining momentum is easier once you have (somehow) arrived near the hilltop.

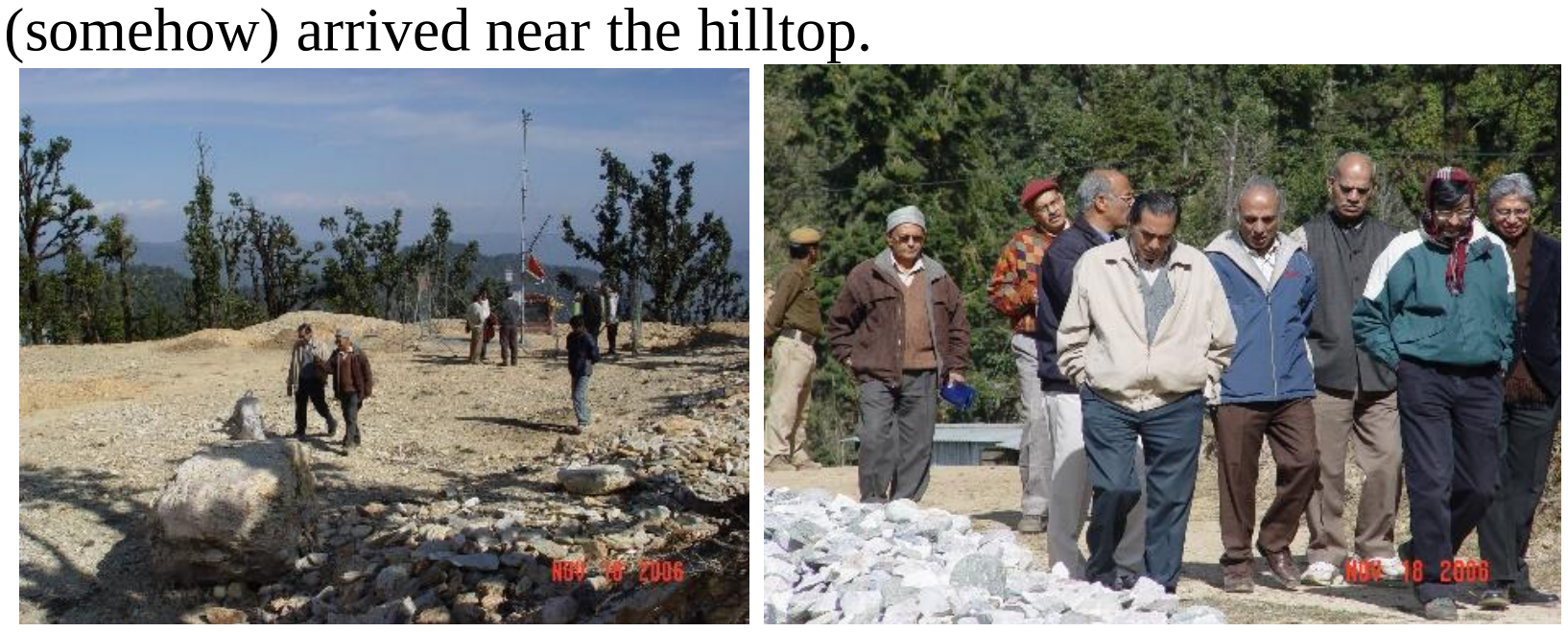

**Fig. 9** On 18 November 2006, late Shri S. C. Tapde and others inspected the location of the 3.6-m DOT (left panel) and the track for the approach road to Devasthal peak (right panel).

It gives us immense pleasure to dedicate this work to the memory of Shri Suresh Chand Tapde (see Fig. 9). Throughout his professional life he remained a pillar of strength in the efforts to boost the indigenous component of any telescope project in which he was substantially involved. While we astronomers are prone to keeping our eyes on stars, he kept his feet on the ground. The saga began with the construction of the 2.34-m VBT by the IIA during the 1970-80s, with Shri Tapde as its Project Manager. VBT, a brainchild of the venerable Prof. M. K. Vainu Bappu, ranked among the large size telescopes, going by the prevailing international standards in those times. Unfortunately, building upon that vital early experience did not progress far, quite unlike the case of other branches of astronomy pursued in India (radio, ultraviolet and X-ray and γ-ray astronomy) where a steady growth in indigenous telescope building capabilities was witnessed. This differential is rather pinching, given that optical astronomy is not only the centre- piece of multi-waveband astronomy, but it is also a branch in which front-line (even Nobel prize winning) research continues to be possible even using small-to-medium size telescopes. This point was analytically emphasized in an article by Gopal-Krishna and Barve (1998) and also appreciated by some of India's thought leaders and planners, like Prof. K. Kasturirangan, as seen from his letter dated 26-December-1998 (Annexure IV). Fortunately, some concrete corrective steps in the domain of Indian optical astronomy have been taken in recent years ( see, e.g. Anupama et al. 2022; Asthana 2022; Section 5).

Our slippages in building upon early success is not limited to science, but also pervades the arena of industry. This systemic lacuna of our society was highlighted by none other than Dr. Homi Bhabha, in the context of India's steel industry. He wrote:

"The steel industry had existed in India since the First World War, and one of the two steel plants was among the largest in the British Commonwealth in the early twenties. Yet when these steel plants had to be expanded, it was necessary to draw upon foreign consultants and engineering firms to plan and carry out the expansions. When the government decided to establish a steel plant at Bhilai the same course was followed, this time with Russian technical collaboration. The third public sector steel plant at Durgapur had similarly to be set up with the help of British consortium…. Thus, the construction and operation of a number of steel plants had not automatically generated the ability to design and build new steel plants." (quoted from 'Bhabha and his Magnificent Obsessions', by G. Venkataraman, University Press, 1994-2008).

This systemic fault-line largely stemmed from the fact that while our steel industry generated steel and plenty of profits, it invested too little in R & D and thus steadily fell behind its international peers in terms of technological advancement and competitiveness, which is so critical for expansion and modernisation of the industry.

The message is writ large.

**Acknowledgements**

We thank an anonymous referee for the constructive suggestions on the original manuscript. The activities chronicled in this article have been enriched by contributions from so many colleagues, however it is not possible to thank them individually due to the space limitation. In establishing the Devasthal observatory, the unstinted support received from late Dr. R. S. Tolia, former Chief Secretary of Uttarakhand; Shri Prabhat C. Chaturvedi, Secretary UP government; late Mr S. C. Tapde, Profs. S. K. Ghosh, P. C. Agrawal, S. N. Tandon, S. Ananthakrishnan and G. Govil; late Dr. A.K. Pandey and Drs. Wahab Uddin, Vijay Mohan, B. B. Sanwal and Brijesh Kumar is thankfully acknowledged. Prof. G. Srinivasan made valuable contribution to the design and construction of the 3.6-m DOT dome and buildings. It is also a pleasure to acknowledge the guidance and support extended by late Prof. S. K. Joshi, chairman, GC, ARIES and the TIFR directors Prof. S.S. Jha and Prof. V. Singh. RS thanks the National Academy of Sciences, India for an Honorary Scientist position; Alexander von-Humboldt Foundation, Germany for the Research Linkage Program and the Director, IIA for hosting. GK is thankful to Indian National Sciences Academy for a Senior Scientist position.

**Annexure I. Documents showing genesis of the TIFR-UPSO collaboration**

TATA INSTITUTE OF FUNDAMENTAL RESEARCH

Ref:TFR/PF/594

April 25, 1996

Dear Prof. Gopal Krishna and Prof. Rengarajan,

As you may be aware, the DAE working Group has pruned down the IXth Plan allocation of TIFR Plan proposals to Rs.250.00 crores as against an outlay of Rs. 330.00 crores projected by the Institute. The TIFR Internal Working Group, at its third meeting held on April 23, 1996 has accordingly pruned down the allocation for the various programmes. For the proposal on the New Otical Telescope the tentative allocation approved is Rs.1800.00 lakhs. This, again, is subject to review and allocation by the Planning Commission.

You are requested to discuss the above Proposal in the respective Group/Facility Committee in the light of the new outlay projected by the Internal working Group and intimate the allocation, with yearwise break up for the proposal. Detailed minutes of the meeting should be provided with justification for support of the proposal. This information will be used by the Internal Working Group for the next round of discussion. It will be appreciated if the information is sent to my office by May 15, 1996.

With regards,

Yours sincerely,

(G.Govil)
Chairman, Budget & Planning Committee

Prof. Gopal Krishna
NCRA, Pune

Prof.T.N. Rengarajan
Space Physics Group
TIFR

---

TATA INSTITUTE OF FUNDAMENTAL RESEARCH

National Centre of the Government of India for Nuclear Science and Mathematics

HOMI BHABHA ROAD, BOMBAY-400 005, INDIA.

Professor VIRENDRA SINGH
DIRECTOR

Telephone : 215 2971, 215 2979
Telegrams : ZETTEKIS
Telex : 011-83009
Voice Mail/
Telefax : 091-22-2152280
Email : vsingh@tifrvax.tifr.res.in

August 27, 1996

Prof. Ram Sagar
Director
Uttar Pradesh State Observatory (UPSO)
Manora Park, Naini Tal (U.P.)

Dear Prof. Ram Sagar,

In its 2nd (1996) meeting held on April 11, 1996 the Physics Faculty of TIFR discussed a joint proposal from the astronomers and astrophysicists of TIFR and its Pune constituent unit, the National Centre for Radio Astrophysics (NCRA/TIFR) to set up a 2-metre class optical telescope in the Central Himalayas. Such a facility will meet their rapidly growing demand for optical observations. The copy of the proposal for IXth Five Year proposal is enclosed.

It was brought to my notice that after an extensive site survey carried out by your observatory during the eighties, a good site has been identified at Devasthal close to your observatory. In view of this and considering the long background of your observatory's involvement in optical astronomy, our institute would like to explore the possibility of setting up a 2-metre class optical telescope at Devasthal, as a collaborative project between our institute and UPSO, on cost-sharing basis (as part of the IXth Five Year Plan). The details of the proposed collaboration would be worked out once the financial commitments for the IXth Five Year proposal becomes clearer.

I look forward to hear your response on this suggestion.

With kind regards,

Yours sincerely,

( V. Singh )

---

DEPARTMENT OF SCIENCE & TECHNOLOGY
GOVERNMENT OF U.P.
LUCKNOW - 226 001
Dated:Jan. 27, 1997
111(PS)557/97

Dear Prof. Singh,

It was a pleasure to visit your esteemed scientific institution. The meeting with you and other scientists on January 2, 1997 was extremely useful. After this visit and the meeting, I am convinced that the proposal of setting up a 2-metre class telescope at Devasthal by TIFR, Mumbai and U.P. State Observatory, Nainital is in the overall national interest and should be strongly supported and funded. The proposed telescope alongwith modern focal plane instruments will contribute significantly to the growth of optical astronomy in the country. As we agreed in the meeting, U.P. Government will meet 50% of the estimated cost ofthe project. Hope your meeting with the Planning Commission went well.

Looking forward to hear from you,

with regards

Yours sincerely,

( P. C. Chaturvedi )

Prof. V. Singh,
Director,
Tata Institute of Fundamental Research,
Homi Bhabha Road,
Mumbai-400 005.

---

TATA INSTITUTE OF FUNDAMENTAL RESEARCH
Homi Bhabha Road; Mumbai 400 005

May 7, 1997

**Presentation and Discussion of Major 9th Plan Programmes**

The Internal Working Group for the 9th Plan Programmes has decided to have presentation and discussion of some of the major research programmes and facilities proposed for the 9th Plan of the Institute. Each presentation, not exceeding 45 minutes duration, will be followed by discussion of upto 30 minutes. Members of the Institute are invited to participate in this programme.

All presentations will take place in the Lecture Theatre .

The Schedule of the programme is as follows :-

| Date | Time | Title | Speaker |
|---|---|---|---|
| Wednesday May 14, 1997 | 1600-1645 | A modern 3 metre class optical telescope in Central Himalayas | Prof. Gopalkrishna |
| | 1645-1715 | Discussion of the above proposal | |
| | 1715-1730 | Tea in West Canteen | |
| Friday May 23, 1997 | 1430-1515 | Superconducting LINAC booster for 14 UD Pelletron accelerator | Prof. M.B. Kurup |
| | 1515-1545 | Discussion of the above proposal | |
| | 1545-1600 | Break | |
| | 1600-1645 | National Centre for Applicable Mathematics | Prof. P.N. Srikanth |
| | 1645-1715 | Discussion of the above proposal | |
| | 1715-1730 | Tea in West Canteen | |
| Wednesday May 28, 1997 | 1430-1515 | National Facility for high field NMR | Prof. R.V. Hosur |
| | 1515-1545 | Discussion of the above proposal | |
| | 1545-1600 | Break | |
| | 1600-1645 | Study of p-p collisions and search for new particles with the CMS detector at LHC (CERN) | Prof. A. Gurtu and Prof. N.K. Mondal |
| | 1645-1715 | Discussion of the above proposal | |
| | 1715-1730 | Tea in West Canteen | |

( K.P. Balakrishnan )
Secretary, Internal Working Group

---

TATA INSTITUTE OF FUNDAMENTAL RESEARCH
Homi Bhabha Road, Mumbai - 400 005

Ref: TFR/Accts/129(98)/ 2635

July 8, 1997

The Internal Working Group for IXth Plan Programmes has made the following budgetary allocation to your Plan programme for the year 1997-98.

(Rupees in lakhs)

| Group Committee | IXth Plan Programme Code | Title of the Programme | Eight Plan Code & amount allotted to cover spill over of 1996-97 | Additional amount allotted for current year 1997-98 | Total amount allotted | Items to be acquired |
|---|---|---|---|---|---|---|
| Common Research Facility | 9P-1001 | Modern Optical Telescope | Nil | 10.00 | 10.00 | A Telescope of 14" or other suitable size with a CCD and cooler and a PC/AT computer for data logging. Also for travel to UPSO, Nainital/Devasthal site. This allocation is only for site survey and preliminary feasibility study of the project. |

I am further directed to request you to note that :

1. Allocated funds are to be used only for the items for which they are approved.
2. Computers and peripherals falling under "Capital Equipments" can be bought from Plan funds only if specific approval is given for these items.
3. Plan funds should not be used for national/international travel unless the funds are approved specifically for this purpose.

S. A. BAVADEKAR
CHIEF ACCOUNTS OFFICER

Prof. Gopal Krishna/~~Prof. S.K. Ghosh~~ NCRA, Pune / Space Physics Group
~~HR Group~~

---

TATA INSTITUTE OF FUNDAMENTAL RESEARCH

Ref:TFR/PF/9plan/ URGENT August 7, 1997

Dear Prof.Gopal Krishna,

Dr.R.Chidambaram, Chairman, AEC, vide his letter No: 31/1/97-R&D-II/847 dated July 25, 1997 addressed to the Director (copy of which was earlier sent to you), has requested to prepare detailed project report in respect of each of the IXth Plan projects of the Institute, in the format prescribed. The project reports are to be placed before the TIFR Council of Management for discussion and approval at its next meeting.

As instructed by the Director, I request you to prepare the report in respect of your Plan-proposal (A Modern 2 metre-class Optical Telescope ...9P-1001) in the format enclosed. Kindly restrict the total amount to the figure communicated to you already by the Internal Working Group.

Kindly submit your report as early as possible, but not later than August 18, 1997, positively.

I solicit your co-operation in the matter.

With regards,

Yours sincerely,

(K.P.Balakrishnan)
Secretary, Internal Working Group,
IXth Plan Programmes

Encl: as above

Prof.Gopal Krishna
NCRA
Pune 411 007

cc: Director- for information

**Annexure II. PSC formed on May 16, 2000 as per MoU signed between TIFR and UPSO**

उत्तर प्रदेश शासन
विज्ञान एवं प्रौद्योगिकी विभाग
सं0:927/45-वि0/2000-4(8)वि0/99
लखनऊ: दिनांक : मई 10, 2000

कार्यालय-ज्ञाप

'देवस्थल नैनीताल में 3 मीटर आप्टिकल दूरबीन की स्थापना' परियोजना के सम्बन्ध में उ0 प्र0 राजकीय वेधशाला, नैनीताल एवं टाटा इन्स्टीट्यूट आफ फण्डामेन्टल रिसर्च (टी0आई0आफ0आर0), मुम्बई के मध्य सम्पन्न हुए समझौता ज्ञापन (मेमोरेण्डम आफ अण्डरस्टैन्डिंग) के प्रस्तर-2.2 के अनुक्रम में परियोजना के सुचारू संचालन हेतु एक प्रोजेक्ट स्टीयरिंग कमेटी का गठन निम्नवत किया जाता है:-

| | | |
|---|---|---|
| 1- | निदेशक,टी0आई0आफ0आर0, मुम्बई, या उनका प्रतिनिधि । | अध्यक्ष |
| 2- | मुख्य सचिव, उत्तर प्रदेश शासन । | सहअध्यक्ष |
| 3- | सचिव, विज्ञान एवं प्रौद्योगिकी विभाग, उ0प्र0 शासन । | सदस्य |
| 4- | संयुक्त सचिव(आर0 एण्ड डी0)आणविक ऊर्जा विभाग, भारत सरकार, नई दिल्ली । | सदस्य |
| 5- | प्रो0 आर0पी0 वर्मा, टी0आई0एफ0आर0, मुम्बई । | सदस्य |
| 6- | प्रो0 गोपाल कृष्णा,टी0आई0एफ0आर0, मुम्बई । | सदस्य |
| 7- | डा0 एस0डी0 सिन्हल, भू0पू0 निदेशक, उ0प्र0 राजकीय वेधशाला, नैनीताल । | सदस्य |
| 8- | प्रो0 नितीश कुमार सान्याल, कुलपति,इन्दिरा गाँधी राष्ट्रीय मुक्त विश्वविद्यालय, इलाहाबाद । | सदस्य |
| 9- | प्रमुख सचिव/सचिव, वित्त विभाग,उ0प्र0 शासन या उनका प्रतिनिधि जो विशेष सचिव स्तर से कम का न हो । | सदस्य |
| 10- | प्रमुख सचिव/सचिव, नियोजन विभाग, उ0प्र0 शासन या उनका प्रतिनिधि जो विशेष सचिव स्तर से कम का न हो । | सदस्य |
| 11- | रजिस्ट्रार, टी0आई0एफ0आर0, मुम्बई । | सदस्य |
| 12- | निदेशक, उ0प्र0 राजकीय वेधशाला, नैनीताल । | सदस्य-संयोजक |

(2) उक्त कमेटी की बैठक वर्ष में कम से कम दो बार अथवा आवश्यकतानुसार दो से अधिक बार अध्यक्ष तथा सहअध्यक्ष की अनुमति से प्रोजेक्ट स्थल, नैनीताल में आहूत होगी ।

योगेन्द्र नारायण
मुख्य सचिव ।

संख्या:927(1)/45-वि0/2000, तद्दिनांक:

प्रतिलिपि प्रोजेक्ट स्टीयरिंग कमेटी के समस्त सदस्यों को सूचनार्थ एवं आवश्यक कार्यवाही हेतु प्रेषित ।

आज्ञा से,
( गंगा राम कुटार )
अनु सचिव ।

**Annexure III. Release of funds by U.P. Government order dated March 27, 199**

प्रेषक

श्री यतीन्द्र मोहन,
अनु सचिव,
उत्तर प्रदेश शासन ।

सेवा में

निदेशक,
उत्तर प्रदेश राजकीय वेधशाला,
नैनीताल ।

[वि]ज्ञान एवं प्रौद्योगिकी विभाग लखनऊ : दिनांक 27 मार्च 1999

विषय : "देवस्थल" नैनीताल में तीन मीटर दूरबीन की स्थापना हेतु वित्तीय स्वीकृति ।

महोदय,

उपर्युक्त विषयक आपके पत्र संख्या 0/2503/1-1 दिनांक 07 दिसम्बर 1998 के सन्दर्भ में मुझे आपसे यह कहने का निदेश हुआ है कि राज्यपाल महोदय वर्तमान वित्तीय वर्ष 1998-99 में देवस्थल (नैनीताल) नामक स्थान पर नई प्रेक्षण सुविधाओं के अन्तर्गत 3 मीटर दूरबीन की स्थापना के प्रयोजनार्थ अधोलिखित विवरण के अनुसार रु0 3,66,00,000-00 (केवल तीन करोड़ छियासठ लाख रुपये) की धनराशि के व्यय किये जाने की स्वीकृति प्रदान करते हैं :-

| | |
|---|---|
| 25- लघु निर्माण कार्य | रु0 11-00 लाख |
| 26- मशीनें और सज्जा/उपकरण और संयंत्र | रु0 355-00 लाख |
| योग | रु0 366-00 लाख |

शर्त यह होगी कि धनराशि कुमायूँ विश्वविद्यालय, नैनीताल के पी0एल0ए0 में रखी जायेगी तथा पी0एल0ए0 से आहरण शासन की अनुमति से ही किया जायेगा ।

2- उपर्युक्त व्यय वर्तमान वित्तीय वर्ष 1998-99 के आय-व्ययक के अनुदान संख्या 70 के लेखा शीर्षक "5425-अन्य वैज्ञानिक तथा पर्यावरणीय अनुसन्धान पर पूँजीगत परिव्यय-आयोजनागत-800-अन्य व्यय-03-उत्तर प्रदेश राजकीय वेधशाला, नैनीताल की देवस्थल परियोजना" के अन्तर्गत सुसंगत इकाईयों के नामे डाला जायेगा ।

3- उपर्युक्त आदेश वित्त विभाग के अशासकीय संख्या ई-11-764/दस-99 दिनांक 25 मार्च 1999 में प्राप्त उनकी सहमति से जारी किये जा रहे हैं ।

भवदीय,
(यतीन्द्र मोहन)
अनु सचिव

संख्या 403(1)/45-वि-99-4(12)वि/97-टी0सी0 तद्दिनांकित

प्रतिलिपि निम्नलिखित को सूचनार्थ एवं आवश्यक कार्यवाही हेतु प्रेषित :-

1- महालेखाकार (प्रथम), उत्तर प्रदेश, इलाहाबाद
2- कोषाधिकारी, नैनीताल
3- वित्त आय-व्ययक अनुभाग 1/2
4- वित्त अधिकारी, कुमायूँ विश्वविद्यालय, नैनीताल
5- वित्त व्यय नियंत्रण अनुभाग-11
6- नियोजन अनुभाग-4
7- निदेशक, टाटा इन्स्टीट्यूट आफ फन्डामेन्टल रिसर्च, होमी भाभा रोड,मुम्बई-400 005

आज्ञा से
(यतीन्द्र मोहन)
अनु सचिव

**Annexure IV. Prof. K. Kasturirangan's letter dated 26 December 1998**

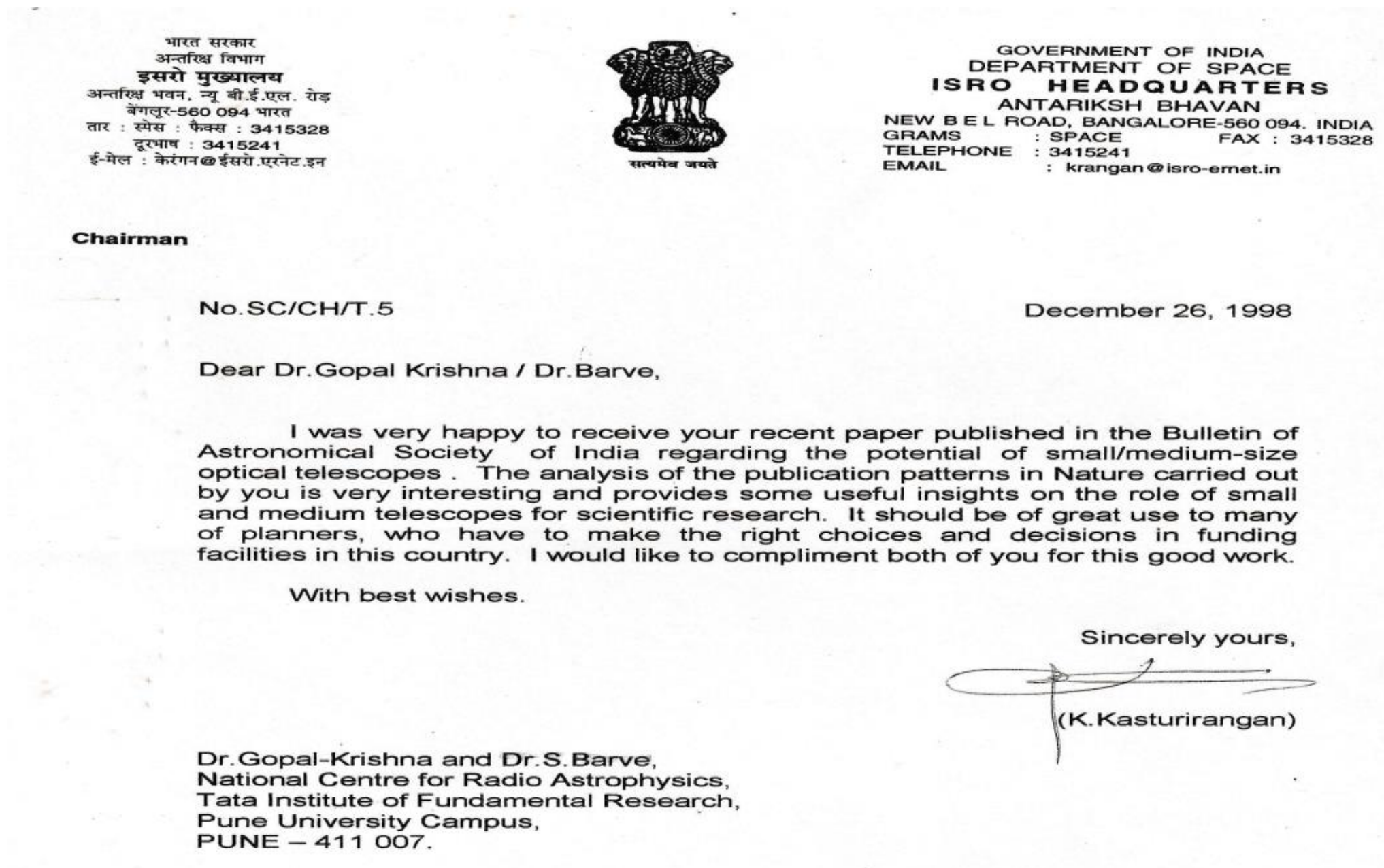

भारत सरकार
अन्तरिक्ष विभाग
**इसरो मुख्यालय**
अन्तरिक्ष भवन, न्यू बी.ई.एल. रोड
बेंगलुर-560 094 भारत
तार : स्पेस : फैक्स : 3415328
दूरभाष : 3415241
ई-मेल : केरंगन@ईसरो.एरनेट.इन

सत्यमेव जयते

GOVERNMENT OF INDIA
DEPARTMENT OF SPACE
**ISRO HEADQUARTERS**
ANTARIKSH BHAVAN
NEW B E L ROAD, BANGALORE-560 094, INDIA
GRAMS : SPACE FAX : 3415328
TELEPHONE : 3415241
EMAIL : krangan@isro-ernet.in

**Chairman**

No.SC/CH/T.5 December 26, 1998

Dear Dr.Gopal Krishna / Dr.Barve,

I was very happy to receive your recent paper published in the Bulletin of Astronomical Society of India regarding the potential of small/medium-size optical telescopes . The analysis of the publication patterns in Nature carried out by you is very interesting and provides some useful insights on the role of small and medium telescopes for scientific research. It should be of great use to many of planners, who have to make the right choices and decisions in funding facilities in this country. I would like to compliment both of you for this good work.

With best wishes.

Sincerely yours,

(K.Kasturirangan)

Dr.Gopal-Krishna and Dr.S.Barve,
National Centre for Radio Astrophysics,
Tata Institute of Fundamental Research,
Pune University Campus,
PUNE – 411 007.

## References

Agrawal, P. C. (2017). AstroSat: From Inception to Realization and Launch. *Journal of Astrophysics and Astronomy,* 38**,** Art. 27. doi:10.1007/s12036-017-9449-6.

Ailawadhi, B., Dastidar, R. et al. (2023). Photometric and spectroscopic analysis of the Type II SN 2020jfo with a short plateau. *MNRAS,* 519, 248-270. doi:10.1093/mnras/stac3234.

Anupama, G. C. et al. (2022). A 10-m class national large optical-IR telescope. *Journal of Astrophysics and Astronomy,* 43, Art. 32. doi:10.1007/s12036-022-09819-6.

Asthana, P. (2022). Mega Science Programme in India – evolution and prospects. *Current Science,* 123, 26–36. https://doi.org/10.18520/cs/v123/i1/26-36.

Bhattacharyya, J.C., Rajan, K.T. (1992). Vainu Bappu Telescope. *Bulletin of the Astronomical Society of India,* 20, 319-343.

Chand, H. et al. (2018). Probing the central engine and environment of AGN using ARIES 1.3-m and 3.6-m telescopes. *Bulletin de la Societe Royale des Sciences de Liege,* 87, 291-298. doi:10.25518/0037-9565.7727.

Dhiman V. et al. (2023). Multiband optical variability of the TeV blazar PG 1553 + 113 in 2019. *MNRAS,* 519, 2796-2811. doi:10.1093/mnras/stac3709.

Dimple, P. et al. (2023). Characterization of a deep-depletion 4K × 4K charge-coupled device detector system designed for ARIES Devasthal faint object spectrograph. *Journal of Astronomical Telescopes, Instruments, and Systems,* 9, 18002; doi:10.1117/1.JATIS.9.1.018002.

Jaiswal, S., Omar, A. (2013). Devasthal Fast Optical Telescope Observations of Wolf-Rayet Dwarf Galaxy Mrk 996. *Journal of Astrophysics and Astronomy*, 34**,** 247-257. doi:10.1007/s12036-013-9182-8.

Gopal-Krishna, Barve S. (1998). Discovery potential of small/medium-size optical telescopes: A study of publication patterns in NATURE (1993-95). *Bulletin of the Astronomical Society of India,* 26, 417-424.

Gopal-Krishna, Wiita P. (2018). Optical monitoring of Active Galactic Nuclei from ARIES. *Bulletin de la Societe Royale des Sciences de Liege,* 87, 281-290. doi:10.25518/0037-9565.7718.

Gopal-Krishna et al. (1995). Intranight optical variability in optically selected QSOs. *MNRAS*, 274, 701-710. doi:10.1093/mnras/274.3.701.

Gopal-Krishna et al. (2003) Clear Evidence for Intranight Optical Variability in Radio-quiet Quasars. *Astrophysical J. Letter,* 586, L25-L28. doi:10.1086/374655.

Gopal-Krishna et al. (2023). Intranight optical variability of low-mass active galactic nuclei: a pointer to blazar-like activity. *MNRASL,* 518, L13-L18. doi:10.1093/mnrasl/slac125.

Gupta, A.K. et al. (2016). Multiband optical variability of three TeV blazars on diverse time-scales. *MNRAS*, 458, 1127-1137. doi:10.1093/mnras/stw377.

Gupta, R. et al. (2022). Photometric studies on the host galaxies of gamma-ray bursts using 3.6m Devasthal optical telescope. *Journal of Astrophysics and Astronomy*, 43, Art 82. doi:10.1007/s12036-022-09865-0.

Joshi, S., De Cat, P. (2019). Overview of the BINA Activities. *Bulletin de la Societe Royale des Sciences de Liege*, 88, 19-30. doi:10.25518/0037-9565.8625.

Joshi, Y. C. et al. (2022). ARIES 130-cm Devasthal Fast Optical Telescope — Operation and Outcome. *Journal of Astronomical Instrumentation,* 11, id 2240004. doi:10.1142/S2251171722400049.

Kumar, A. et al. (2022). Photometric calibrations and characterization of the 4K×4K CCD imager, the first-light axial port instrument for the 3.6m DOT. *Journal of Astrophysics and Astronomy,* 43, Art 27. doi:10.1007/s12036-022-09798-8.

Kumar, B. et al. (2018). 3.6-m Devasthal Optical Telescope Project: Completion and first results. *Bulletin de la Societe Royale des Sciences de Liege,* 87, 29–41. doi:10.25518/0037-9565.7454.

Kumar, B. et al. (2022). First Light Preparations of the 4m ILMT. *Journal of Astronomical Instrumentation,* 11, id 2240003. doi:10.1142/S2251171722400037.

Lata, S. et al. (2021). Photometric observations of NGC 281: detection of 228 variable stars. *MNRAS*, 504, 101-117. doi:10.1093/mnras/stab885.

Maurya, J. et al. (2021). Photometric and Kinematic Study of the Open Clusters SAI 44 and SAI 45. *Astron. J.***,** 162, id 64, doi:10.3847/1538-3881/ac0138.

Maurya, J. et al. (2023). Investigating Stellar Variability in the Open Cluster Region NGC 381. *Astron. J.***,** 165**,** id. 90, doi:10.3847/1538-3881/acad7e.

Mohan, V. et al. (1999). Atmospheric extinction at Devasthal, Naini Tal. *Bulletin of the Astronomical Society of India*, 27, 601-608.

Mishra, S. et al. (2021). A search for blazar activity in broad-absorption-line quasars. *MNRASL*, 507**,** L46-L51. doi:10.1093/mnrasl/slab095.

Ningombam, S. S. et al. (2021). Evaluation of fractional clear sky over potential astronomical sites. *MNRAS*, 507, 3745-3760. doi:10.1093/mnras/stab1971.

Ojha, D.K. et al. (2018). Prospects for star formation studies with infrared instruments (TIRCAM2 and TANSPEC) on the 3.6-m Devasthal Optical Telescope. *Bulletin de la Societe Royale des Sciences de Liege*, 87, 58-67. doi:10.25518/0037-9565.7480.

Ojha, V. et al. (2020). Comparative intranight optical variability of X-ray and $\gamma$-ray-detected narrow-line Seyfert 1 galaxies. *MNRAS*, 493, 3642-3655. doi:10.1093/mnras/staa408.

Omar, A. et al. (2017). Scientific capabilities and advantages of the 3.6 meter optical telescope at Devasthal, Uttarakhand. *Current Science*, 113, 682–685. doi:10.18520/cs/v113/i04/682-685.

Omar, A. et al. (2019a). First-light images from low dispersion spectrograph-cum-imager on 3.6-meter Devasthal Optical Telescope. *Current Science*, 116, 1472-1478. doi:10.18520/cs/v116/i9/1472-1478.

Omar, A. et al. (2019b). Optical detection of a GMRT-detected candidate high-redshift radio galaxy with 3.6-m Devasthal optical telescope. *Journal of Astrophysics and Astronomy*, 40, Art 9. doi:10.1007/s12036-019-9583-4.

Pandey, R. et al. (2020). Stellar Cores in the Sh 2-305 H II Region. *Astrophysical J.*, 891, id. 81. doi:10.3847/1538-4357/ab6dc7.

Pandey, S.B. et al. (2021). Photometric, polarimetric, and spectroscopic studies of the luminous, slow-decaying Type Ib SN 2012au. *MNRAS*, 507**,** 1229-1253. doi:10.1093/mnras/stab1889.

Panwar, N. et al. (2017). Low-mass young stellar population and star formation history of the cluster IC 1805 in the W4 H II region. *MNRAS*, 468**,** 2684-2698. doi:10.1093/mnras/stx616.

Panwar, N. et al. (2022). Deep V and I CCD photometry of young star cluster NGC 1893 with the 3.6m DOT. *Journal of Astrophysics and Astronomy*, 43, Art 7. doi:10.1007/s12036-021-09785-5.

Priyatikanto, R. et al. (2023). Characterization of Timau National Observatory using limited in situ measurements. *MNRAS*, 518**,** 4073-4083. doi:10.1093/mnras/stac3349.

Raiteri, C.M. et al. (2021). The dual nature of blazar fast variability: Space and ground observations of S5 0716+714. *NRAS*, 501, 1100-1115. doi:10.1093/mnras/staa3561.

Sagar, R. (2022). History of ARIES: a premier research institute in the area of observational sciences. *Indian Journal of History of Sciences,* 57**,** 227-247. doi:10.1007/s43539-022-00054-0.

Sagar, R. (2023). Indo-Belgian co-operation in Astrophysics: From inception to future prospects. *Bulletin de la Societe Royale des Sciences de Liege*, 92 (accepted).

Sagar, R. et al. (2000). Evaluation of Devasthal site for optical astronomical observations. *Astronomy & Astrophysics Supplement Series*, 144, 349–362. doi:10.1051/aas:2000213.

Sagar, R. et al. (2011). The new 130-cm optical telescope at Devasthal, Nainital. *Current Science*, 101, 1020–1023.

Sagar, R., Kumar, B., Omar, A. (2019). The 3.6 metre Devasthal Optical Telescope: From inception to realization. *Current Science,* 117, 365–381. doi:10.18520/cs/v117/i3/365-381.

Sagar, R., Kumar, B., Sharma, S. (2020). Observations with the 3.6 meter Devasthal Optical Telescope. *Journal of Astrophysics and Astronomy,*41, Art.33. doi:10.1007/s12036-020-09652-9.

Sagar, R. et al. (2022). Optical Observations of star clusters NGC 1513 and NGC 4147; white dwarf WD1145+017 and K band imaging of star forming region Sh2–61 with the 3.6 meter Devasthal Optical Telescope. *Journal of Astrophysics and Astronomy*, 43, Art 31, doi:10.1007/s12036-022-09815-w.

Sanwal, B.B. et al. (2018). History of initial ffty years of ARIES: A Major National Indian Facility for Optical Observations. *Bulletin de la Societe Royale des Sciences de Liege*, 87, 15–28. doi:10.25518/0037-9565.7445.

Sharma. S. et al. (2022a). TANSPEC: TIFR-ARIES Near-infrared Spectrometer. *Pub. Astron. Soc. Pacific*, 134, id.085002. doi:10.1088/1538-3873/ac81eb.

Sharma, S. et al. (2022b). First Lunar Occultation Results with the TIRCAM2 Near-Infrared Imager at the Devasthal 3.6-m Telescope. *Journal of Astronomical Instrumentation,* 11, id. 2240002. doi:10.1142/S2251171722400025.

Sharma, S. et al. (2023). Teutsch 76: a Deep Near-Infrared Study. *Journal of Astrophysics and Astronomy*, 44, id 46. DOI: 10.1007/s12036-023-09936-w.

Sicardy, B. et al. (2021). Pluto's Atmosphere in Plateau Phase Since 2015 from a Stellar Occultation at Devasthal. *Astrophysical J. Letter*, 923, L31. doi:10.3847/2041-8213/ac4249.

Stalin, C. S. et al. (2001). Seeing and microthermal measurements near Devasthal top. *Bulletin of the Astronomical Society of India*, 29, 39–52.

Surdej, J., Hickson, P., Kumar, B., Misra, K. (2022). First Light with the 4-m International Liquid Mirror Telescope. *Physics News*, 52, 25-28.

Surdej, J. et al. (2023) The 4m International Liquid Mirror Telescope project: preliminary results. *Bulletin de la Societe Royale des Sciences de Liege*, 92 (accepted).

Swarup, G. (2021) The Journey of a Radio Astronomer: Growth of Radio Astronomy in India. *Annual Review of Astronomy and Astrophysics*, 59:1-19. doi: 10.1146/annurev-astro-090120-014030.

Troja, E et al. (2022) A nearby long gamma-ray burst from a merger of compact objects. *Nature*, 612, 228-231. doi:10.1038/s41586-022-05327-3.